\documentclass[preprint]{aastex}
\usepackage{amsmath}
\usepackage{amssymb}
\usepackage{amstext}
\usepackage{graphicx}
\usepackage{epsfig}

\usepackage{color}
\definecolor{myColor}{rgb}{0.9,0.9,0.9}

\begin{document}

\title{Millisecond Dips in Sco X-1 are Likely the Result of High-Energy Particle Events}

\author{T. A. Jones\altaffilmark{1}, A. M. Levine \altaffilmark{2}, E. H. Morgan \altaffilmark{2}, and S. Rappaport\altaffilmark{1}}   

\altaffiltext{1}{Department of Physics and Kavli Institute for Astrophysics and Space Research, MIT, Cambridge, MA 02139; {\tt }}

\altaffiltext{2}{Kavli Institute for Astrophysics and Space Research, MIT, Cambridge, MA 02139; {\tt }}


\begin{abstract}

Chang et al.~(2006) reported millisecond duration dips in the X-ray
intensity of Sco X-1 and attributed them to occultations of the source
by small trans-Neptunian objects (TNOs). We have found evidence that
these dips are in fact not astronomical in origin, but rather the
result of high-energy charged particle events in the RXTE PCA
detectors.

\end{abstract}


\section{Introduction}
\label{sec:intro}

Chang et al.~(2006) found statistically significant one to two
millisecond duration dips in the count rate during X-ray observations
of Sco X-1 carried out with the Proportional Counter Array (PCA) on
the Rossi X-ray Timing Explorer (RXTE) and attributed them to
occultations of the source by small objects orbiting the Sun beyond
the orbit of Neptune, i.e., trans-Neptunian objects (TNOs).  In all,
Chang et al.~(2006) found some 58 dips in approximately 322 ks of Sco
X-1 observations.  Given that the RXTE spacecraft moves through the
diffraction-widened shadows of any TNOs at a velocity of $\sim30$ km
s$^{-1}$, dips of $\sim$2 ms duration should correspond to a TNO size
of $\sim$60 m.  If the identification of these dips with occultations
by TNOs is correct, the dips would provide extremely valuable
information on the number and distribution of solar system objects of
$\sim20$-100 m in size.

\section{Average Properties of Dips in the Sco X-1 Count Rates}
\label{sec:data}

Subsequent to the report by Chang et al.~(2006), we identified
$\gtrsim$200 dips of the type they describe in some 500 ks of RXTE/PCA
observations of Sco X-1.  In an attempt to detect the small counting
rate {\em increases} that one might expect from diffraction sidelobes
if these are indeed occultation events, we formed an average dip
profile by superposing the PCA light curves that include the
dips. This was done after fitting a Gaussian to each dip in order to
estimate its centroid time, maximum depth, and width.  We use the full
width at half maximum (FWHM) of the fitted Gaussian as our estimate of
dip width.  Then, using the fitted centroid time and dip width, we
shift and stretch each dip to a common centroid time and width before
accomplishing the superposition.  We expected to see sidelobes with
intensity $\sim$5\% greater than the mean count rate determined
substantially away from the superposed dips.  We find no evidence for
diffraction sidelobes in the superposed light curves, despite having
statistics sufficient to reduce fluctuations to $\sim$1\% (1~$\sigma$)
of the mean count rate.

Three additional problems with the occultation interpretation are
manifest from the dip profiles.  First, the summed dip profile is
distinctly asymmetric in shape (as Chang et al. suggested for many of
the individual dips).  Second, the distribution of dip widths is
narrower than what one would expect from occultations by bodies with a
power-law size distribution of index -4, i.e., there are fewer than
expected statistically significant dips with Gaussian FWHM widths
greater than $\sim2$ ms.  Third, when diffraction effects are taken
into consideration, the majority of dip widths are shorter than the
minimum widths one would expect from occultations by bodies at
distances of $\sim 40$ AU regardless of their size distribution.

\begin{figure*}   
\epsscale{0.85}
\plotone{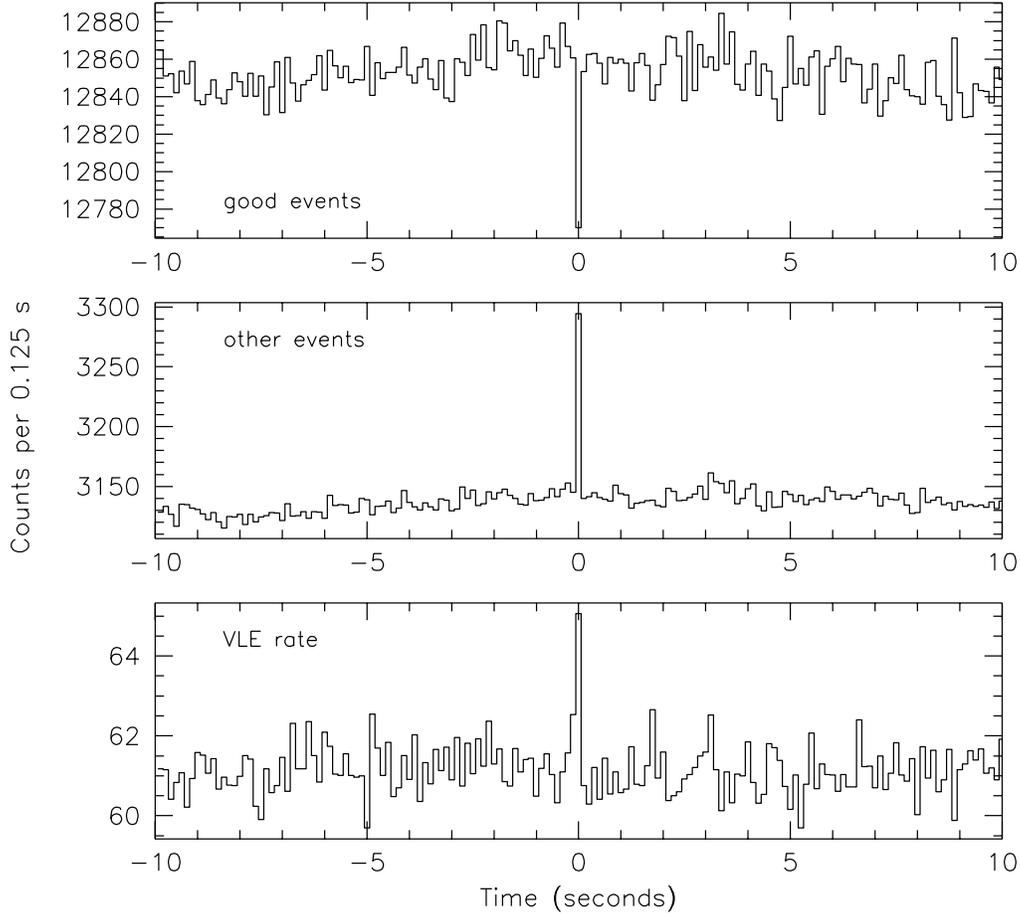}

\caption{Counts per 1/8-s time bin of different types of PCA detector
events superposed, i.e., averaged, around the times of 201 dips.  The
superposition was accomplished such that the bin at $time = 0$~s
includes the identified dips.  The counts include events from all
(typically 3 to 5) of the operating PCUs. Top panel: good xenon
counting rate data. The decrease in counting rate due to the dips is
apparent.  The small ($\sim$0.6\%) drop in the counting rate is
explained in the text.  Middle panel: counting rate data of events
that are not good events, propane-only events, or VLEs.  This category
includes multiple-LLD events (see text). A highly significant
enhancement in the vicinity of the dips is evident.  Bottom panel: VLE
event rate data superposed around the dip times, also showing a
statistically significant peak.  Note that the peak in the VLE event
rate is approximately one VLE event per detector per dip event (i.e.,
$\sim$4 excess events per dip).}

\label{fig:xraylc}
\end{figure*}
  
\section{Search for an Alternate Explanation}
\label{sec:expl}

These findings prompted us to further explore alternative explanations
for the dips. Only one hypothesis appeared to be worthwhile to pursue,
viz., that the dips are caused by electronic dead time in response to
some type of charged particle shower in the spacecraft \citep[see][and
references therein, for technical information on the PCA]{pca06}.
Coincidences within a $\sim10$~$\mu$s window among two or more of the
measurement chains in a Proportional Counter Unit (PCU) are used to
identify charged particle events.  However, the intensity of Sco X-1
is so high that there is a substantial rate of coincidences due to the
detection of two X-ray photons within the 10~$\mu$s window.  For some
observations, the rates of such so-called two-LLD events were
telemetered with millisecond time resolution; the dips are apparent in
these data as expected.  In contrast, no information is available on
the non-X-ray background during the Sco X-1 observations with
millisecond time resolution.  Counts of good events, very large events
(VLEs), propane-layer events, and a catch-all category of other types
of events that includes multiple LLD events are available
at 1/8 s time resolution from Standard Mode 1; most other types of
data are only available with 16-s time resolution.

The VLE flag for a PCU is set when the electronics detects an event in
that PCU with energy greater than $\sim$100 keV; this can happen in
response to the ionization produced by a single charged particle or to
that produced by multiple charged particles which penetrate the
detector nearly simultaneously. Such a large event can produce ringing
in the front-end of an electronic measurement chain.  Therefore, in
response to the occurrence of a VLE, the digital logic shuts down the
electronic processing of events in that PCU for a fixed time period,
chosen to be 50 $\mu$s for almost all of these Sco X-1
observations. In addition, each of the 6 main xenon layer measuring
chains is disabled until its charge drops to an acceptable level. In
order for the detector to be shut down for an extended period ($>$
1~ms), an extraordinary amount of charge must be deposited on most of
the 6 main measuring chains; it is unclear, at present, whether this
can happen in response to a single charged particle.

The prime purpose of the propane layer is to distinguish events caused
by soft ($\sim 1$ MeV) electrons from those caused by X-rays. However,
Sco X-1 is such a strong source that during observations of Sco X-1
the bulk of the events seen only in the propane layer are due to
X-rays from the source.  If the source flux is diminished due to an
intervening TNO, we would expect that the propane rate would decrease
similarly to the xenon rate.

Figure 1 shows counts of three different types of events in 1/8-s time
bins superposed around the times of 201 dips.  In each panel, the
centers of the short ($\sim$2 ms) dips have been placed in the bin at
$time = 0$.  The top panel shows the rates of good events, i.e., those
not identified as being due to charged particles, in the main xenon
layers of all operating PCUs, and clearly shows the superposed dips;
two-LLD events are not included in these rates.  The counting rate
drops by only $\sim$0.8\% because of the dilution of a $\sim$2 ms dip
within a 128 ms wide bin.  By contrast, the middle panel shows the
{\em enhancement} of the counting rate in Standard Mode 1 ``other''
events\footnote{These were incorrectly identified as propane-only
events in the first version of this paper.} in the vicinity of the
dips.  The peak is highly significant ($\sim$38 $\sigma$).  The bottom
panel corresponds to the VLE event rate superposed around the dip
times.  This peak is also statistically very significant ($\sim$8
$\sigma$).  The increase in the VLE rate is more or less consistent
with the detection of $\sim$1 VLE per PCU per dip.

The enhancements in the other event and VLE rates around the times of
the dips indicate that there is an increase in the rate of detection
of non-X-ray events.  We speculate that these non-X-ray events
interrupt normal event processing for 1-2 milliseconds in all or most
of the PCUs roughly once per hour due to the collection of very large
amounts of charge. Such an energetic event may be the consequence of a
particle shower produced by the collision of a high-energy cosmic ray
with a nucleus in the RXTE spacecraft. In any case, further
clarification of the causes of the observed dips would be of interest.

\section{Conclusions}
\label{sec:conc}

While our results cast doubt on whether any true occultation events
have been detected, one cannot yet conclude that no such events have
been detected. We intend to conduct further investigations of the dip
phenomenon and its possible causes, and we will work to obtain a new
measurement of, or upper limit on, the rate of occurrence of
occultations of Sco X-1.


\begin{thebibliography}{}

\bibitem[Chang et al.(2006)]{chang06} Chang, H.-K., King, 
S.-K., Liang, J.-S., Wu, P.-S., Lin, L.~C.-C., \& Chiu, J.-L.\ 2006, \nat, 
442, 660 


\bibitem[Jahoda et al.(2006)]{pca06} Jahoda, K., Markwardt, 
C.~B., Radeva, Y., Rots, A.~H., Stark, M.~J., Swank, J.~H., Strohmayer, 
T.~E., \& Zhang, W.\ 2006, \apjs, 163, 401 

\end{thebibliography}
\end{document}